\documentclass[a4paper,11pt]{article}
\usepackage{pos}

\title{Finite-size scaling of Lee-Yang zeros and its application to the 3-state Potts model and heavy-quark QCD}
\ShortTitle{Finite-size scaling of Lee-Yang zeros -- application to Potts model and heavy-quark QCD}

\author*[a]{Tatsuya Wada}
\author[a,b]{Masakiyo Kitazawa}
\author[c]{Kazuyuki Kanaya}

\affiliation[a]{Yukawa Institute for Theoretical Physics,
Kyoto University, Kyoto 606-8502, Japan}
\affiliation[b]{
  J-PARC Branch, KEK Theory Center, 
  Institute of Particle and Nuclear Studies, KEK, \\ Tokai, Ibaraki 319-1106, Japan}
\affiliation[c]{Tomonaga Center for the History of the Universe, University of Tsukuba, \\ Tsukuba, Ibaraki 305-8571, Japan}

\emailAdd{tatsuya.wada@yukawa.kyoto-u.ac.jp}
\emailAdd{kitazawa@yukawa.kyoto-u.ac.jp}
\emailAdd{kanaya@ccs.tsukuba.ac.jp}

\abstract{
We propose a new general method to study critical points (CP) 
using the finite-size scaling of Lee-Yang zeros (LYZ)~\cite{Wada:2024qsk}. 
We first study the LYZ in the three-dimensional Ising model on finite lattices.
We show that the ratios of multiple LYZ (Lee-Yang-zero ratios: LYZR) have useful scaling properties similar to the Binder cumulants, providing us with a novel method to study CP. 
In numerical simulations of the Ising model, we confirm that this method works well.
We then apply the method to analyze the CP in the three-dimensional three-state Potts model and finite-temperature QCD in heavy-quark region, which are believed to belong to the same universality class as the Ising model. 
In these models, the partition function at complex parameters can be evaluated by the reweighting method, which allows us to determine the LYZ by varying coupling parameters continuously around the CP.
We demonstrate that the LYZR method is powerful in determining the location of the CP in these models.
}

\FullConference{The 41st International Symposium on Lattice Field Theory (LATTICE2024)\\
 28 July - 3 August 2024\\
Liverpool, UK\\}

\begin{document}
\begin{flushright}
YITP-25-12, J-PARC-TH-0311, UTHEP-797
\end{flushright}
\maketitle

\section{Introduction}
%Clarification of 
Clarifying the QCD phase diagram as a function of temperature $T$ and baryon chemical potential $\mu_B$ is one of the most %important issues 
interesting subjects in physics.
%In particular, 
Many researchers are paying attention to the QCD critical point (QCD-CP) because if such a CP exists, thermodynamic properties in its vicinity receive tight constraints through the universal scaling laws.
% At the hadronic scale, QCD is inherently non-perturbative, preventing the use of perturbative calculations and leaving lattice QCD as the only ab initio method available.
Monte Carlo simulations of QCD at non-vanishing $\mu_B$, however, are plagued by the sign problem, 
%To avoid it, various methods including Taylor expansion, imaginary chemical potential, reweighting, complex Langevin, and Lefschetz thimble methods, have been developed~\cite{?}. 
%Despite these efforts, 
%However, 
and a precise determination of the location of QCD-CP is not yet achieved.

Recently, a novel approach for determining the QCD-CP has been proposed based on the Lee-Yang edge singularity (LYES)~\cite{Kortman:1971zz, %} was proposed to determine the QCD-CP~\cite{
Stephanov:2006dn,Ejiri:2014oka,An:2017brc,Basar:2021gyi,Rennecke:2022ohx,Johnson:2022cqv,Singh:2023bog,Karsch:2023rfb,Dimopoulos:2021vrk,Basar:2023nkp,Zambello:2023ptp,Clarke:2024ugt,Skokov:2024fac}, which is the end point of the distribution of the Lee-Yang zeros (LYZ) in the infinite-volume limit.
In Refs.~\cite{Clarke:2024ugt,Basar:2023nkp}, the LYZ on the complex baryon chemical potential have been investigated using the QCD simulations at imaginary chemical potential and the Pad\'{e} approximation. Then, identifying %are the zeros of the partition function in the complex coupling parameter space.
%Identifying 
the first LYZ %obtained by QCD simulation at imaginary chemical potentials 
with the LYES in the infinite-volume limit, the scaling relation of the LYES has been applied to extract the location of the CP.
Although this method will work well when the system volume is %with 
sufficiently large,
%system volumes, 
influence of finite-size effects arising from the difference between the first LYZ and LYES should be evaluated carefully in practical simulations. 

%This method is based on the Lee-Yang zeros formalism. 
%Lee-Yang zeros (LYZ) are the zeros of the partition function in the complex coupling parameter space~\cite{???}.
%T.D.~Lee and C.N.~Yang showed that the LYZ are related to the order of phase transition. 
%From the scaling relation of LYES, we can expect the location of CP. Some groups demonstrate this method in $2+1$-flavor QCD using the imaginary chemical potential method and the Pade approximation and they show the location of the QCD-CP from the Lattice QCD data at physical quark mass value.

%While their method is only valid in the large spatial volume regarded as the infinite volume, lattice calculation is performed in only finite volume so an alternative method with finite-size effect is necessary. One of the methods including the finite-size effects is finite-size scaling (FSS) analysis.
%Few researchers have addressed to the problem of the FSS of LYZ, so we try to formulate a novel method to determine the CP with FSS.

In this report, we develop a new method for studying CP using the finite-size scaling (FSS) properties of LYZ~\cite{Wada:2024qsk}.
We first study the three-dimensional Ising model at finite volumes and show %. We find 
that ratios of multiple LYZ (Lee-Yang zero ratios: LYZR) have a useful FSS property similar to that of the Binder cumulant~\cite{Binder:1981sa},
providing us with an alternative method, the LYZR method, to study the CP.
We also discuss the extension of the method for studying CP in general systems in the same universality class. 
We numerically show that the method works well in the three-dimensional three-state Potts model and QCD in heavy-quark region.
%Furthermore, we verify this method in the 3d-3state Potts model and Heavy-Quark QCD.
%Finally, we summarize our findings and discuss future prospects for this method.

\section{Finite-size scaling of LYZ in Ising model}

\subsection{Lee-Yang-zero ratio}
Let us start with the three-dimensional Ising model with the Hamiltonian
\begin{align}
    H(h)=-J\sum_{\langle ij \rangle} s_is_j-h\sum_{i} s_i, \label{eq:def_Ising_Hamiltonian}
\end{align}
on the square lattice of size $L^3$ with the external magnetic field $h$, where $\sum_{\langle i,j\rangle}$ represents the summation over all pairs of adjacent sites. 
In the infinite-volume limit $L\to \infty$, the Ising model is known to have a CP at temperature $T=T_c$ and $h=0$, where the critical temperature $T_c=4.51152322(1)$ has been determined precisely in Monte Carlo simulations~\cite{Ferrenberg:2018zst}. In what follows, we set $J=1$ to make $T$ and $h$ dimensionless. 

We denote the partition function at finite $L$ and the reduced temperature $t=(T-T_c)/T_c$ as $Z(t,h,L^{-1})={\rm Tr}e^{-H/T}$. 
The LYZ are the zeros of $Z(t,h,L^{-1})$ on the complex-$h$ plane for real $t$.
It is known that the LYZ are located on the imaginary axis with $\textrm{Re}h =0$~\cite{Lee:1952ig}.
Because the Ising model has the $Z(2)$ symmetry,
$Z(t,-h,L^{-1})=Z(t,h,L^{-1})$,
the LYZ in this model
%and 
have a reflection symmetry: 
if $h=h_0$ is a LYZ, $h=-h_0$ is also a LYZ.
We denote the LYZ with $\textrm{Im} h>0$ as $h=h^{(n)}_\textrm{LY}(t,L)$, where $n=1,2,\cdots$ is labeled such that $\textrm{Im} h^{(n)}_{\textrm{LY}}$ increases with increasing $n$. 
Since $Z(t,h,L^{-1})$ is a regular function of $t$ and $h$ for finite $L$, $h^{(n)}_\textrm{LY}(t,L)$ are also regular in $t$ for finite $L$.

According to the FSS, $Z(t,h,L^{-1})$ in the vicinity of the CP is represented by the scaling function $\tilde{Z}$ as 
\begin{align}
    Z(t,h,L^{-1}) = \tilde{Z}(L^{y_t}t,L^{y_h}h) ,
\label{eq:FSS}
\end{align}
for sufficiently large $L$,
where the scaling dimensions are %given by 
$y_t\simeq 1.588$ and $y_h\simeq 2.482$ for the three-dimensional Ising model~\cite{Ferrenberg:2018zst}. %, whose values have been analyzed with high precision.
Since the LYZ are zeros of Eq.~\eqref{eq:FSS}, %the LYZ obey the FSS 
they obey
\begin{align}
    h_{\rm LY}^{(n)}(t,L) = L^{-y_h} \, \tilde{h}_{\rm LY}^{(n)}(L^{y_t}t) ,
    \label{eq:tildeh_LY}
\end{align}
with $\tilde{h}_{\rm LY}^{(n)}(\tilde t)$ satisfying $\tilde Z(\tilde t,\tilde{h}_{\rm LY}^{(n)}(\tilde t))=0$.

At $t<0$, corresponding to the discontinuity of the first-order phase transition at $h=0$, the LYZ are distributed densely around $h=0$ in the infinite-volume limit~\cite{Lee:1952ig}.
In fact, it is easily shown that the LYZ %here 
are aligned on the imaginary $h$ axis with an equal separation as
\begin{align}
    h_{\rm LY}^{(n)}(t,L)=a(t)(2n-1)/L^3 \; \xrightarrow[L\to\infty]{} \; 0 
    \quad
    (t<0,\;n:\textrm{finite}),
    \label{eq:LYES0}
\end{align} 
for sufficiently large $L$, where $a(t)$ is a pure-imaginary function~\cite{Ejiri:2005ts}.

At $t>0$, since $\lim_{L\to\infty}Z(t,h,L^{-1})$ is a regular function at $h=0$, the distributions of LYZ for $L\to\infty$ must terminate at pure-imaginary points $h=\pm h_{\rm LYES}(t)$ away from the real axis of $h$, which are called the LYES~\cite{Kortman:1971zz}. This implies that
\begin{align}
    h_{\rm LY}^{(n)}(t,L) \;
    %=L^{-y_h} {\tilde h}_{\rm LY}^{(n)}(L^{y_t}t) 
    \xrightarrow[L\to\infty]{} \; h_{\rm LYES}(t) 
    \quad
    (t>0,\;n:\textrm{finite}).
    \label{eq:LYES}
\end{align} 
Since the right-hand side of Eq.~\eqref{eq:LYES} does not depend on $L$, only possible asymptotic behavior of ${\tilde h}_{\rm LY}^{(n)}({\tilde t})$ %defined in Eq.~\eqref{eq:tildeh_LY} 
for $\tilde t\to\infty$ is ${\tilde h}_{\rm LY}^{(n)}({\tilde t})\propto {\tilde t}^{y_h/y_t}$, which yields $h_{\rm LYES}(t)\propto t^{y_h/y_t}$~\cite{BENA_2005}.

\begin{figure}
    \centering
\includegraphics[width=0.25\textwidth]{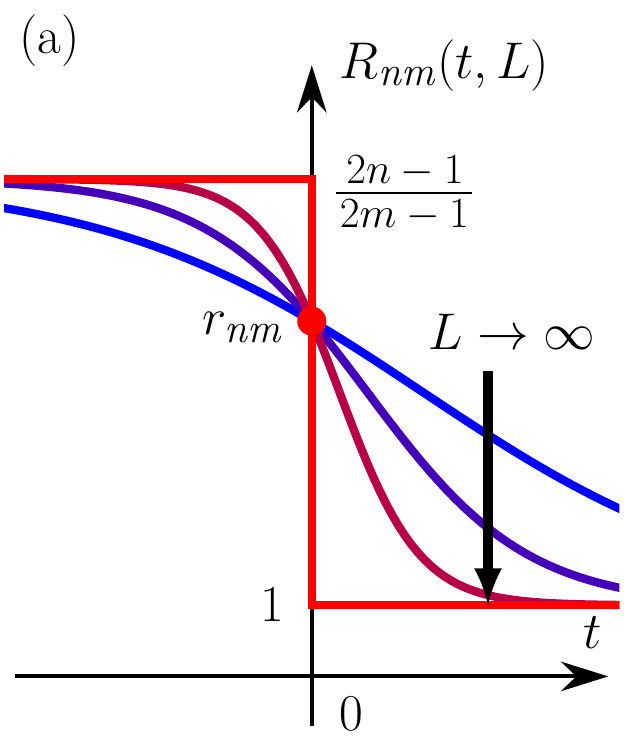}
\hspace{0.15\textwidth}
\includegraphics[width=0.25\textwidth]{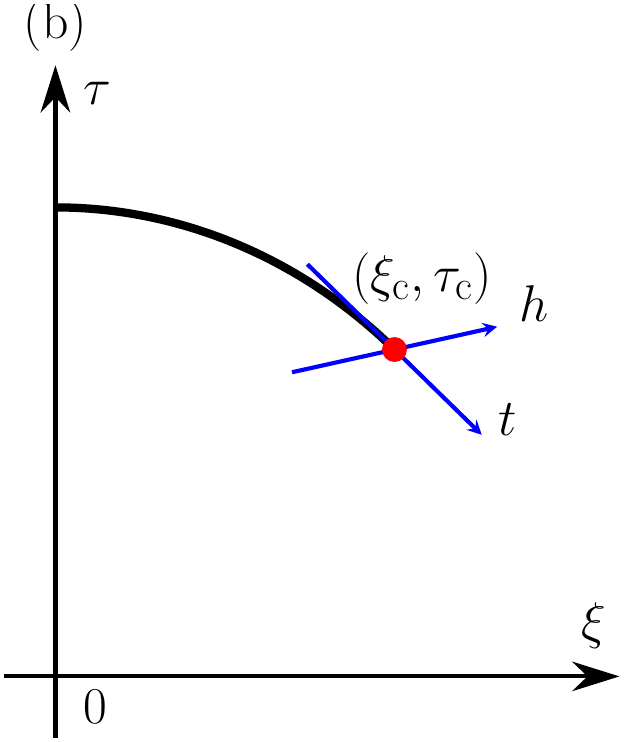}
    \caption{(a) Schematic plot of Lee-Yang zero ratio $R_{nm}(t,L)$ as function of $t$~\cite{Wada:2024qsk}. Blue, purple, and brown curves represent $R_{nm}(t,L)$ at small, medium, and large $L$'s, respectively. In the limit $L\to\infty$, $R_{nm}(t,L)$ approach the red step function. % at the CP $t=0$.
    (b) Schematic phase diagram of the three-dimensional three-state Potts model~\eqref{eq:H_Potts}. The black-solid line shows the first-order phase transition that terminates at a CP denoted by the red-filled circle. The Ising variables $t$ and $h$ encoded through the mapping of the scaling function to this system are shown by the blue arrows.}
    \label{fig:cartoon}
\end{figure}

Now we focus on the ratio of two LYZ at the same $L$~\cite{Wada:2024qsk},
\begin{align}
    R_{nm}(t,L) = \frac{h_{\rm LY}^{(n)}(t,L)}{h_{\rm LY}^{(m)}(t,L)}
    = \frac{\tilde h_{\rm LY}^{(n)}(L^{y_t}t)}{\tilde h_{\rm LY}^{(m)}(L^{y_t}t)} .
    \label{eq:Rnm}
\end{align}
From Eqs.~\eqref{eq:LYES0} and \eqref{eq:LYES}, one easily finds that Eq.~\eqref{eq:Rnm} behaves in the $L\to\infty$ limit %leads to
as
\begin{align}
    R_{nm}(t) \xrightarrow[L\to\infty]{}
    \begin{cases}
        \frac{2n-1}{2m-1} & (t<0) \\
        1 & (t>0)
    \end{cases}
    \qquad
    (\mbox{finite}~ n,m),
    \label{eq:Rlim}
\end{align}
as shown by the red step function in Fig.~\ref{fig:cartoon}~(a). Equation~\eqref{eq:Rlim} suggests that the behavior of LYZR~\eqref{eq:Rnm} changes rapidly at $t=0$. This behavior may be useful in locating the CP in numerical simulations. 

To gain a better insight into the behavior of Eq.~\eqref{eq:Rnm} around $t=0$, we Taylor expand $\tilde{h}_{\rm LY}^{(n)}(\tilde t)$ at $\tilde t=0$ as 
\begin{align}
    \tilde{h}_{\rm LY}^{(n)}(\tilde t) = i\big( X_{n} + Y_{n}\tilde t + {\cal O}(\tilde{t}\,^2) \big) ,
    \label{eq:XY} 
\end{align}
with real coefficients $X_n$ and $Y_n$, which is allowed from the regularity of $\tilde{h}_{\rm LY}^{(n)}(\tilde t)$.
Substituting this into Eq.~\eqref{eq:Rnm}, one obtains 
\begin{align}
    & R_{nm}(t,L) = r_{nm} + c_{nm} L^{y_t} t + {\cal O}(t^2) ,
    \label{eq:Rlinear}
\end{align}
with $r_{nm} = X_n/X_m$ and $c_{nm} = r_{nm}\,(Y_n/X_n - Y_m/X_m )$. 
Equation~\eqref{eq:Rlinear} means that $R_{nm}(0,L)=r_{nm}$ is independent of $L$, while the slope at $t=0$ scales as $L^{y_t}$. In other words, $R_{nm}(t,L)$ for different $L$ intersect at the CP at $t=0$, as schematically shown in Fig.~\ref{fig:cartoon}~(a). The properties~\eqref{eq:Rlim} and \eqref{eq:Rlinear} are similar to the Binder cumulants~\cite{Binder:1981sa}. They thus provide us with an alternative method to determine the CP from the intersection point in numerical simulations. 

%These properties are expected around the CP when $L$ is sufficiently large.
%When $L$ is not large enough, contamination of regular terms in FSS may introduce $L$-dependent corrections to these properties. 

\subsection{Numerical verification of the LYZR method}

To verify Eq.~\eqref{eq:Rlinear}, we perform Monte Carlo simulations of the three-dimensional Ising model with the Hamiltonian Eq.~\eqref{eq:def_Ising_Hamiltonian}. 
%The critical temperature $T_c=4.51152322(1)$ has been determined with high precision %using the Binder cumulant method
%through the analyses of susceptibilities~\cite{Ferrenberg:2018zst}. 
%In the present study, 
We generate spin configurations using the Wolff algorithm~\cite{Wolff:1988uh} with $L=24$, 48, and 64. 
Simulations are performed near the CP %at $(T,h)=(T_\textrm{sim},0)$ 
with 8 parameters between $T_\textrm{sim}=4.5110$ and $4.5120$ at $h=0$. 
At each simulation point, $2\times10^6$ configurations are generated every ten iterations after thermalization. 
%Thanks to the cluster algorithm, the autocorrelation time is  at most 
%of the order small around 
%$\order{10}$ updates. 

\begin{figure}
    \centering
    \includegraphics[width=0.8\linewidth]{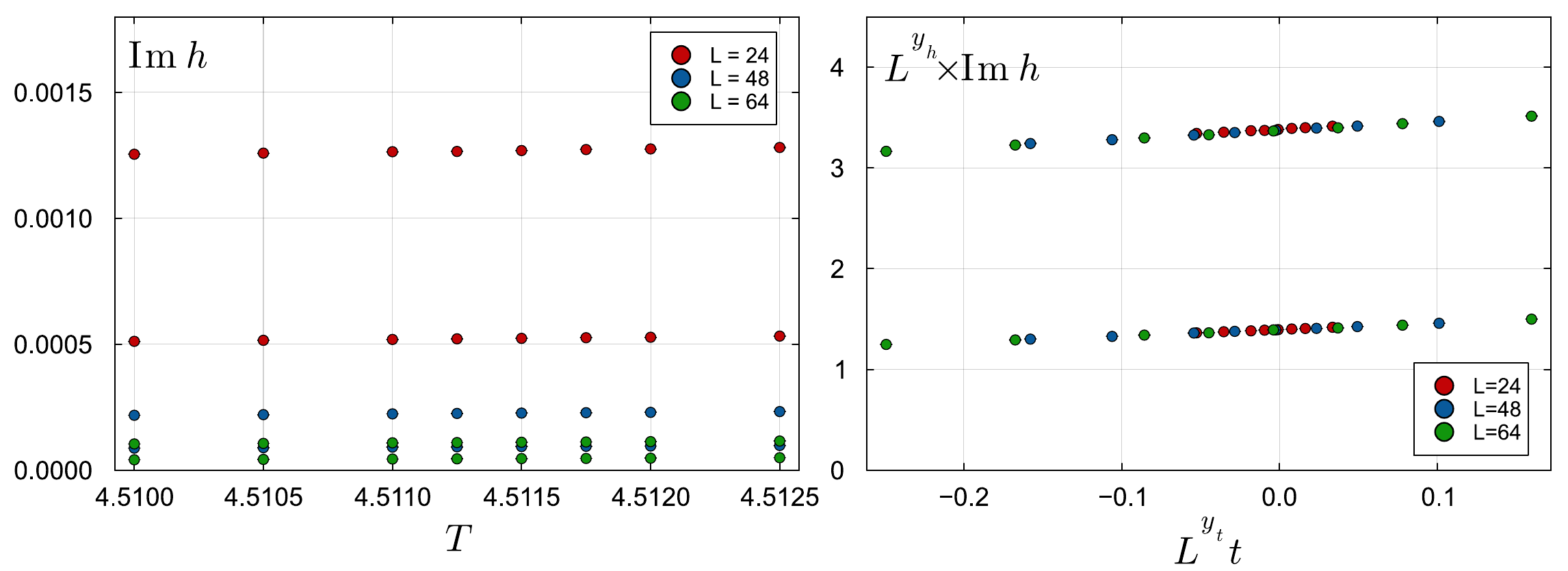}
    \caption{Left: Imaginary part of the LYZ in the three-dimensional Ising model as a function of $T$ for $L=24$, 48, and 64.
    Right: The same data after the rescaling according to Eq.~(\ref{eq:tildeh_LY}).}
    \label{fig:Ising_scaling}
\end{figure}

We determine LYZ for each $L$ using the reweighting method~\cite{PhysRevLett.61.2635,Ejiri:2005ts}.
% i.e. we search for the zeros of 
% \begin{align}
    % \frac{{Z}(T,h,L^{-1})}{{Z}(T,\textrm{Re}h,L^{-1})}
    % = \frac{
    % \big\langle e^{-T^{-1}{H}(h)+{T^{-1}_\textrm{sim}}{H}(0)} \big\rangle_L
    % }{
    % \big\langle e^{-T^{-1}{H}(\textrm{Re}h)+{T^{-1}_\textrm{sim}}{H}(0)} \big\rangle_L
    % } ,
    % \label{eq:Zreweight}
% \end{align}
% with $T\in\mathbb{R}$ and $h\in\mathbb{C}$ for each simulation parameter, where $\langle\cdot\rangle_L$ denotes the average over the configurations at $(T_{\rm sim},0)$ of size $L^3$. The numerical cost to calculate Eq.~\eqref{eq:Zreweight} does not depend on $L$ and is negligibly small compared to that for updates of configurations. 
%Whereas reweighting methods suffer from the overlapping problem when $(T,h)$ is not close to $(T_{\rm sim},0)$, we found that this problem is well suppressed in our analysis, as demonstrated below. 
%The methods yield the following results. First, we obtain the LYZ and 
%We obtain the LYZ for each $L$ and operating finite-size scaling on them, we obtained the result as fig~\ref{fig:Ising_scaling}. The figure on the left shows the raw LYZ, which includes both the first and second LYZ for each volume and demonstrates that as the lattice size increases, the LYZ approaches the real $h$ h-axis. 
We confirm that the overlapping problem of the reweighting method is well suppressed in our analyses.
In the left panel of Fig.~\ref{fig:Ising_scaling}, 
we show the imaginary parts of the first ($n=1$) and second ($n=2$) LYZ as functions of $T$ for three $L$. Statistical errors are smaller than the markers.
We find that these results depend on $L$ sensitively.
%approach the real $h$-axis as $L$ is increased, both in the first order and in the crossover sides 
%The figure on the right shows the LYZ after the scaling operation. It can be seen that the raw 1st and 2nd LYZ each move onto a single line by scaling.
In the right panel of Fig.~\ref{fig:Ising_scaling}, we plot the same result by rearranging the axes according to the FSS of Eq.~(\ref{eq:tildeh_LY}) with %, where we adopt 
the values of $y_t$ and $y_h$ in the Ising model~\cite{Kos:2016ysd}. 
We see that the scattered data of LYZ at different $L$ and $T$ converge well to universal curves, as predicted by Eq.~(\ref{eq:tildeh_LY}).

\begin{figure}
    \centering
    \includegraphics[width=0.8\linewidth]{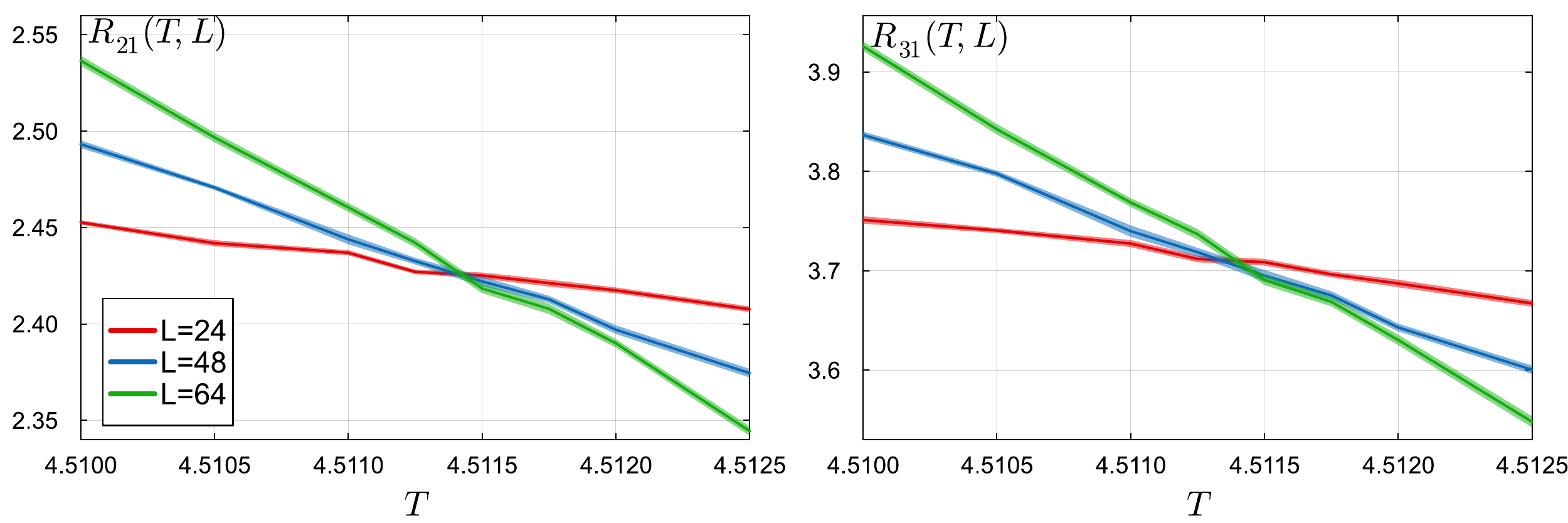}
    \caption{LYZR $R_{21}(t,L)$ (left) and $R_{31}(t,L)$ (right) 
    in the three-dimensional Ising model as functions of $T$ for $L=24$, 48, and 64.
    }
    \label{fig:Ising_R}
\end{figure}

In Fig.~\ref{fig:Ising_R}, we show $R_{21}(t,L)$ and $R_{31}(t,L)$ as functions of $T$ for various $L$. 
The shaded bands show statistical errors estimated by the jackknife method with %the bin size of $20$. 
$20$ bins.
The figure shows that $R_{n1}$ increase (decrease) as $L$ becomes larger at $t<0$ ($t>0$) as expected from Eqs.~\eqref{eq:Rlim} and~\eqref{eq:Rlinear}.
One also finds that $R_{n1}$ for various $L$ intersects approximately at a single point within the errors.
The crossing point is located at $T\simeq4.5114$, which is consistent with
the result of $T_c$ in Ref.~\cite{Ferrenberg:2018zst}.
% $T_c=4.51152322(5)$
This result confirms the validity of the LYZR method in the Ising model.
For further verification, we are now extending our numerical analysis to larger $L$ values~\cite{Wada:2025xx}.

\section{LYZR method for CP in general systems}

\subsection{LYZR in general systems}
Next, we extend the previous argument to deal with CP in general systems that belong to the same universality class as the Ising model. 
For clarity of presentation, we first consider the three-dimensional three-state Potts model with the Hamiltonian
\begin{align}
    \frac{{\cal H}(\tau,\xi)}T = -\tau\sum_{\langle i,j\rangle} \delta_{\sigma_i\sigma_j} - \xi \sum_i \delta_{\sigma_i,1},
    \label{eq:H_Potts}
\end{align}
on the simple cubic lattice of size $L^3$ with the periodic boundary conditions.
Here, $\sigma_i$ takes three states $\sigma_i=1,2,3$.
%with the subscript denoting the lattice site.
%This model partition function ${\cal Z}(\tau,\xi,l^{-1})$ is described by two variables $\tau$ and $\xi$ with lattice size $L$.
The schematic phase diagram of this model is shown in Fig.~\ref{fig:cartoon}~(b).
At $\xi = 0$, this model has a first-order phase transition associated with the spontaneous breaking of the $Z(3)$ symmetry. 
This first-order phase transition %line in the $\tau-\xi$ plane 
is terminated at a CP at $(\tau,\xi)=(\tau_{\rm c},\xi_{\rm c})$ as depicted in Fig.~\ref{fig:cartoon}~(b).
This CP belongs to the Ising universality class.
In the vicinity of the CP and for sufficiently large $L$, the partition function ${\cal Z}(\tau,\xi,L^{-1})$ of the Potts model is related to the partition function $Z(t,h,L^{-1})$ of the Ising model by the linear transformation of the coupling parameters %, $\check t(\tau,\xi)$ and $\check h(\tau,\xi)$: 
\begin{align}
    {\cal Z}\left(\tau,\xi,L^{-1}\right) \propto Z\left(\check t(\tau,\xi),\check h(\tau,\xi),L^{-1}\right) ,
    \label{eq:F=F}
\end{align}
with 
\begin{align}
    \begin{pmatrix}
        \check t \\ \check h
    \end{pmatrix}
    =
    \begin{pmatrix}
        a_{11} & a_{12} \\
        a_{21} & a_{22} 
    \end{pmatrix}
    \begin{pmatrix}
        \tau-\tau_{\rm c} \\ \xi-\xi_{\rm c}
    \end{pmatrix}
    \equiv
    A
    \begin{pmatrix}
        \delta\tau \\ \delta\xi
    \end{pmatrix} ,
    \label{eq:A}
\end{align}
% \textcolor{red}{(KK: $=$ was replaced by $\propto$. Please check!)}
where the $t$ axis encoded onto the $\tau$--$\xi$ plane should be parallel to the first-order phase-transition line at the CP as shown in Fig.~\ref{fig:cartoon}~(b)~\cite{Rehr:1973zz,Wilding:1997xx,Karsch:2000xv}. 

We define the LYZ of the Potts model as zero points of ${\cal Z}\left(\tau,\xi,L^{-1}\right)$ for $\xi\in\mathbb{C}$ and $\tau\in\mathbb{R}$ and 
denote them with ${\rm Im}\xi>0$ by $\xi=\xi_{\rm LY}^{(n)}(\tau,L)$, where the definition of the label $n$ is the same as before.
From Eq.~\eqref{eq:F=F}, we find that 
\begin{align}
    \check h\big(\tau,\xi_{\rm LY}^{(n)}(\tau,L)\big) 
    = L^{-y_h} \, \tilde{h}_{\rm LY}^{(n)}\big(L^{y_t}\check t(\tau, \xi_{\rm LY}^{(n)}(\tau,L)) \big),
    \label{eq:hxi}
\end{align}
where $\tilde{h}_{\rm LY}^{(n)}$ is %the Ising scaling functions 
defined in Eq.~\eqref{eq:tildeh_LY}.
% Equation~\eqref{eq:hxi} together with Eqs.~\eqref{eq:A} and~\eqref{eq:XY} leads to
% \begin{align}
    % &L^{y_h} \big( a_{21} \delta\tau + a_{22} (\xi_{\rm LY}^{(n)}(\tau,L)-\xi_{\rm c}) \big) 
    % = i\big( X_n + Y_n L^{y_t} \big( a_{11} \delta\tau + a_{12} (\xi_{\rm LY}^{(n)}(\tau,L)-\xi_{\rm c}) \big) \big)+ {\cal O}(\delta\tau^2),
% \end{align}
% which gives
% \begin{align}
    % \xi_{\rm LY}^{(n)}(\tau,L)  
    % = \xi_{\rm c} + \frac{ iX_n - ( a_{21} L^{y_h} -iY_n a_{11} L^{y_t} )\delta\tau}{ a_{22} L^{y_h} -iY_n a_{12}L^{y_t} },
    % \label{eq:xi_LY}
% \end{align}
% where the terms of order ${\cal O}(\delta\tau^2)$ are suppressed for simplicity.
% Using $0<y_t<y_h$ and expanding Eq.~\eqref{eq:xi_LY} by $l^{-1}$ one obtains 
% \begin{align}
    % &{\rm Re}\xi_{\rm LY}^{(n)}(\tau,L)  
    % = \xi_{\rm c} - \frac{a_{21}}{a_{22}}\delta\tau + {\cal O}(L^{2\bar y}),
    % \label{eq:Rexi} 
    % \\
    % &{\rm Im}\xi_{\rm LY}^{(n)}(\tau,L)  
    % = \frac{X_n}{a_{22}} L^{-y_h} + \frac{Y_n\det A}{a_{22}^2} L^{\bar y} \delta\tau + {\cal O}(l^{2\bar y}),
    % \label{eq:Imxi}
% \end{align}
% with $\bar y=y_t-y_h<0$.
Then, by combining Eqs.~\eqref{eq:XY}, \eqref{eq:A} and~\eqref{eq:hxi} one finds~\cite{Wada:2024qsk} %that the real and imaginary parts of $\xi_{\rm LY}^{(n)}$ are given by
\begin{align} 
    &{\rm Re}\,\xi_{\rm LY}^{(n)}(\tau,L)
    = \xi_{\rm c} - \frac{a_{21}}{a_{22}}\delta\tau + {\cal O}(L^{2\bar y}), \label{eq:Rexi} 
    \\
    &{\rm Im}\,\xi_{\rm LY}^{(n)}(\tau,L)
    = \frac{X_n}{a_{22}} L^{-y_h} + \frac{Y_n\det A}{a_{22}^2} L^{\bar y} \delta\tau + {\cal O}(L^{2\bar y}), \label{eq:Imxi} 
    \end{align}
with $\bar y = y_t - y_h < 0$.
%Equation~\eqref{eq:Rexi} shows that $(\tau,{\rm Re}\xi_{\rm LY}^{(n)}(\tau,L))$ moves along the $t$-axis with $h=0$ for $L\to\infty$. 
Equations~\eqref{eq:tildeh_LY} and \eqref{eq:Imxi} imply that ${\rm Im}\,\xi_{\rm LY}^{(n)}(\tau,L)\propto \delta\tau^{y_h/y_t}$ for $L\to\infty$ and $\delta\tau\to0$~\cite{Stephanov:2006dn}. 
Finite-size corrections to these results can be estimated by explicitly calculating higher-order terms omitted in Eqs.~\eqref{eq:Rexi} and~\eqref{eq:Imxi}. 

To introduce the LYZR for this case, 
%As LYZR, 
we consider the ratios of ${\rm Im}\,\xi_{\rm LY}^{(n)}(\tau,L)$. 
By expanding them by $\delta\tau$ and $L^{-1}$, one obtains~\cite{Wada:2024qsk}
\begin{align}
    {\cal R}_{nm}(\tau,L) 
    =& \frac{{\rm Im}\,\xi_{\rm LY}^{(n)}(\tau,L)}{{\rm Im}\,\xi_{\rm LY}^{(m)}(\tau,L)}
    % \\\notag 
    = \big( r_{nm} + C_{nm} L^{y_t} \delta\tau + {\cal O}(\delta\tau^2) \big) 
    \times \big( 1 + D_{nm} L^{2\bar y} + {\cal O}(L^{4\bar y})\big),
    \label{eq:Rlinear2}
\end{align}
where $C_{nm}=c_{nm}\det A /a_{22}$ and $D_{nm}= -(Y_n^2-Y_m^2)a_{12}^2/a_{22}^2$. For $L\to\infty$, Eq.~\eqref{eq:Rlinear2} is dominated by the first bracket on the far-right-hand side. 
This means that the intersection point of Eq.~\eqref{eq:Rlinear2} converges to the CP for $L\to\infty$ as in the Ising model. 
Note that $r_{nm}=\lim_{L\to\infty}{\cal R}_{nm}(\tau_{\rm c},L)$ is the same as $r_{nm}$ of Eq.~\eqref{eq:Rlinear}, i.e. the value of ${\cal R}_{nm}(\tau,L)$ at the intersection point is universal. 
However, for finite $L$, the second bracket in Eq.~\eqref{eq:Rlinear2} gives rise to a deviation unless $a_{12}=0$. 
%It is also easy to see that Eq.~\eqref{eq:Rlinear2} obeys Eq.~\eqref{eq:Rlim} in the $L\to\infty$ limit away from $\delta\tau=0$. 

Comparing Eq.~\eqref{eq:Rlinear2} with the corresponding relation for the Binder cumulants~\cite{Binder:1981sa}, one finds that the leading finite-$L$ correction term $D_{nm} L^{2\bar y}$ in the LYZR decreases faster than that %e corresponding term ${\rm const.}L^{\bar y}$ 
of the Binder cumulants as $L$ increases~\cite{Wada:2024qsk}.
This would be an advantage of the LYZR method.

% Now, let us compare Eq.~\eqref{eq:Rlinear2} with the Binder-cumulant method~\cite{Binder:1981sa}. For locating a CP on the $\tau$--$\xi$ plane, one may define the fourth-order Binder cumulant as ${\cal B}_4(\tau,l)={\rm min}_\xi [(\partial^4 {\cal F}(\tau,\xi,l^{-1})/\partial \xi^4)/(\partial^2 {\cal F}(\tau,\xi,L^{-1})/\partial \xi^2)^2]+3$~\cite{Karsch:2000xv,Jin:2017jjp,Kiyohara:2021smr} with the free energy ${\cal F}(\tau,\xi,L^{-1})=-T \ln {\cal Z}(\tau,\xi,L^{-1})$ with temperature $T$. One then obtains~\cite{Jin:2017jjp,Cuteri:2020yke}
% \begin{align}
    % {\cal B}_4(\tau,L) 
    % = \big( b_4 + c_4 L^{y_t} \delta\tau + {\cal O}(\delta\tau^2) \big) 
    % \big( 1 + d_4 L^{\bar y} + {\cal O}(l^{2\bar y})\big),
    % \label{eq:Rlinear3}
% \end{align}
% where $d_4$ is proportional to $a_{12}$. Comparing this result with Eq.~\eqref{eq:Rlinear2}, one finds that the second bracket in Eq.~\eqref{eq:Rlinear3} converges slower than that in Eq.~\eqref{eq:Rlinear2} for $L\to\infty$. This implies that the finite-volume effect from $a_{12}\ne0$ is suppressed more quickly for $L\to\infty$ in ${\cal R}_{nm}(\tau,L)$ than ${\cal B}_4(\tau,L)$, which would be an advantage of the former.
% \section{Numerical analysis}

These arguments are applicable to any system that belongs to the same universality class.
In the following, we examine Eq.~\eqref{eq:Rlinear2} numerically in the Potts model and QCD in the heavy-quark region. %, performing Monte Carlo simulations.

\subsection{Numerical verification of the LYZR method in the Potts model}

We generate configurations of Eq.~\eqref{eq:H_Potts} by the heat-bath algorithm on $L=24$, 30, 40, 50, 60, and 70 lattices at simulation points $(\tau_{\rm sim},\xi_{\rm sim})$ near the CP with $\xi_{\rm sim}=0.0007$, 0.00075, and 0.0008. %, and corresponding $\tau_{\rm sim}$ chosen from the Table~I of Ref.~\cite{Karsch:2000xv}. 
At each simulation point, we perform measurements on $10^6$ configurations separated by ten heat-bath updates after thermalization.
See Ref.~\cite{Wada:2024qsk} for details.
%In Ref.~\cite{Karsch:2000xv}, the CP was determined using the Binder-cumulant method.
%in relation to CP in heavy-quark QCD~\cite{Cuteri:2020yke,Kiyohara:2021smr,Ashikawa:2024njc}.

\begin{figure}
    \centering
    \begin{minipage}{0.66\linewidth}
        \includegraphics[trim=0cm 0cm 21cm 0cm,clip,width=1.0\linewidth]{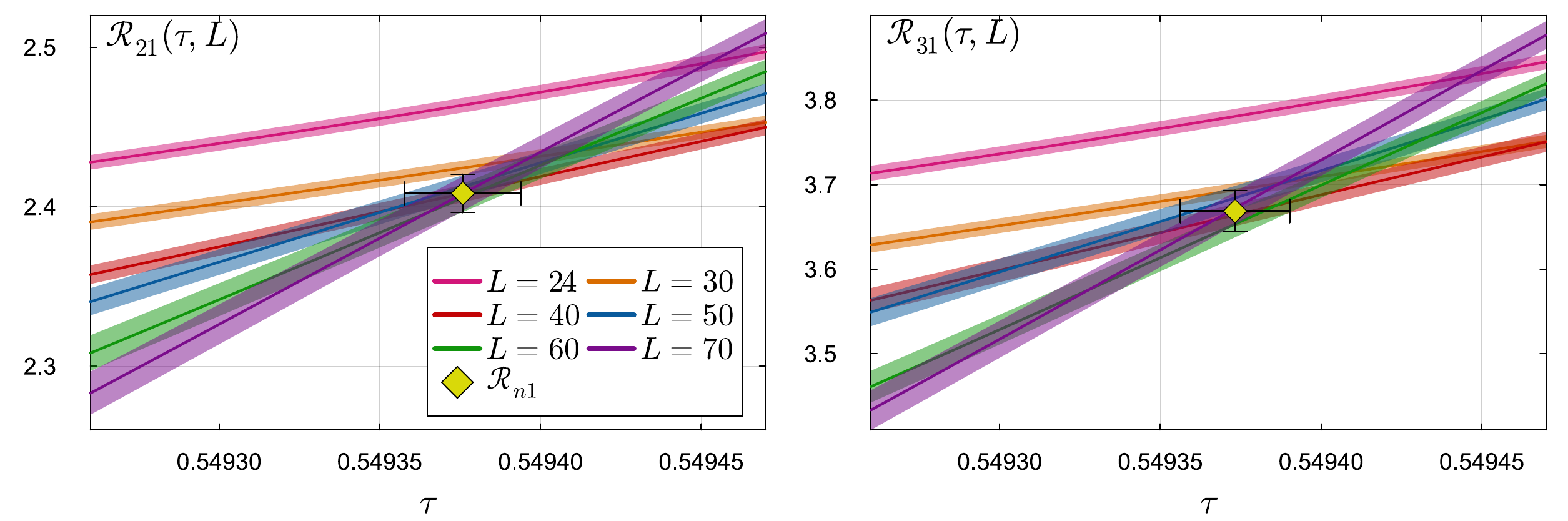}
    \end{minipage}
    \begin{minipage}{0.33\linewidth}
        \includegraphics[width=1.0\linewidth]{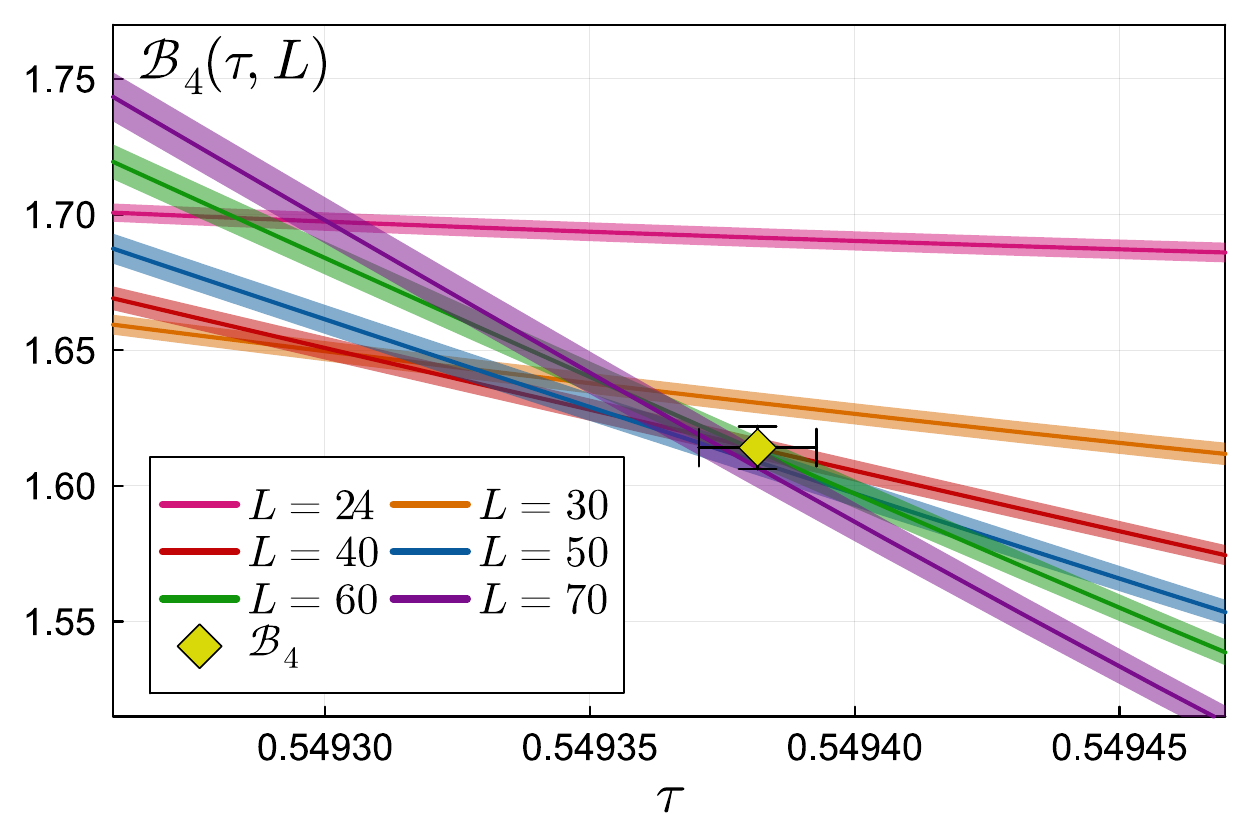}
    \end{minipage}
    \caption{LYZR ${\cal R}_{21}(\tau,L)$ (left) and ${\cal R}_{31}(\tau,L)$ (center), together with the Binder cumulant ${\cal B}_4(\tau,L)$ (right) in the three-dimensional three-state Potts model~\cite{Wada:2024qsk}. The diamonds are the fit results of the intersection point.}
    \label{fig:RB4}
\end{figure}

% \begin{table}[b]
    % \caption{Fit results of the CP parameters and $\chi^2/{\rm dof}$. }
    % \label{tab:result}
    % \centering
    % \begin{tabular}{l|cccc}
    % fit data & $\tau_{\rm c}$ & $y_t$ & $r_{n1}$ or $b_4$ & $\chi^2/\rm{dof}$
    % \\
    % \hline
    % ${\cal R}_{21}$ & 
   % 0.549375(18) & 1.53(19) & 2.408(12) & 0.38 \\
    % ${\cal R}_{31}$ & 
   % 0.549373(17) & 1.66(19) & 3.669(24) & 0.38 \\
    % ${\cal R}_{41}$ & 
   % 0.549372(22) & 1.71(21) & 4.861(36) & 0.55 \\
    % ${\cal R}_{21,31,41}$ & 
   % 0.549379(14) & 1.70(16) &    --      & 0.56 \\
    % ${\cal B}_4$ &
   % 0.549382(11) & 1.63(13) & 1.614(8) & 0.69
    % \end{tabular}
% \end{table}

In Fig.~\ref{fig:RB4}, we show the LYZR ${\cal R}_{21}(\tau,L)$ (left) and ${\cal R}_{31}(\tau,L)$ (center) as functions of $\tau$ and $L$, where the shaded bands represent statistical errors estimated by the jackknife method with $20$ bins. 
The figure shows that ${\cal R}_{n1}(\tau,L)$ for various $L$ intersect at almost a common point for $L\ge40$, while the results for $L=24$ and 30 show clear deviations.

To obtain the critical value $\tau=\tau_{\rm c}$, we perform four-parameter chi-square fits to the data of ${\cal R}_{n1}(\tau,L)$ for $L\ge40$ ($12$ data points in total) with an ansatz ${\cal R}_{n1}(\tau,L)=r+c(\tau-\tau_{\rm c})L^{y_t}$ with $r$, $c$, $\tau_{\rm c}$, and $y_t$ being the fitting parameters. The effects of the second bracket in Eq.~\eqref{eq:Rlinear2} are neglected since no clear deviation from the intersection point is visible for $L\ge40$ in Fig.~\ref{fig:RB4}. 
The fitting results are $\tau_c = 0.549375(18)$ for $\mathcal{R}_{21}$ and $\tau_c = 0.549373(17)$ for $\mathcal{R}_{31}$, as shown by the diamonds in Fig.~\ref{fig:RB4}.
%We also perform an eight-parameter correlated fit to ${\cal R}_{21}(\tau,L)$, ${\cal R}_{31}(\tau,L)$, and ${\cal R}_{41}(\tau,L)$ with common $\tau_{\rm c}$ and $y_t$, to obtain $\tau_c = 0.549379(14)$. See Ref.\cite{Wada:2024qsk} for details.

In the right panel of Fig.~\ref{fig:RB4}, we show the Binder cumulant ${\cal B}_4(\tau,L)$ obtained on the same configurations~\cite{Wada:2024qsk}.
The intersection point obtained from the fit with the same procedure as above, shown by the diamond, is $\tau_c = 0.549382(11)$, which is consistent with Ref.~\cite{Karsch:2000xv}. 
We find that $\tau_{\rm c}$ from the Binder cumulant analysis is consistent with those from the LYZR method.

\subsection{Heavy-Quark QCD}

We now turn to QCD in the heavy-quark region. %, \textit{i.e.}, around the top right corner of the Columbia plot. 
In the heavy-quark limit corresponding to pure gauge $\textrm{SU(3)}$ Yang-Mills theory, the confined and deconfined phases are separated by the first-order phase transition.
%The first-order deconfining transition in the heavy-quark limit (pure gauge $\textrm{SU(3)}$ Yang-Mills theory) 
The first-order transition
weakens as the quark masses decrease 
and eventually terminates at a CP, which has been investigated by the Binder-cumulant method in Refs.~\cite{Cuteri:2020yke,Kiyohara:2021smr,Ashikawa:2024njc}.

Now, let us adopt the LYZR method for a determination of 
%In this study, we determine the location of 
this CP. %in heavy-quark QCD using the LYZR method.
For simplicity, we focus on the case of degenerate $2$-flavor ($N_f=2$) QCD, although generalization to nondegenerate and/or many flavor cases is straightforward. 
We perform the analysis on gauge configurations generated in Ref.~\cite{Kiyohara:2021smr} %perform Monte Carlo simulations 
on four-dimensional Euclidian lattices with temporal lattice size $N_t=4$ and aspect ratios $LT=N_\textrm{s}/N_\textrm{t}=8$, 9, 10, and 12.
These configurations are generated at %We adopt 
the leading-order heavy-quark QCD action of the hopping-parameter expansion (HPE)~\cite{Kiyohara:2021smr,Ashikawa:2024njc}, which reads
\begin{align}
    S_{g+\textrm{LO}} = -6 N_\textrm{t}N_\textrm{s}^3 \beta \hat{P} - 2 N_f N_c \kappa^4 (48N_\textrm{t}N_\textrm{s}^3\hat{P}+32N_\textrm{s}^3\hat{\Omega}_R),
    \label{eq:LO_HQ_action}
\end{align}
for $N_t=4$ with $N_c=3$, where $\beta$ is the inverse gauge coupling and $\kappa=1/(2am_q+4)$ is the hopping parameter with the bare quark mass $m_q$. 
Here, $\hat{P}$ and $\hat{\Omega}$ are the plaquette and Polyakov-loop operators normalized such that they become unity in the weak-coupling limit. 
%We can simulate Eq.~\eqref{eq:LO_HQ_action} 
In Ref.~\cite{Kiyohara:2021smr}, $6\times10^5$ configurations have been generated for three simulation points around the CP 
by the heat-bath algorithm with over-relaxation, working quite efficiently on parallel computers.
%, which is efficiently implemented on parallel computers. % with a total simulation cost comparable to quenched QCD simulations~\cite{Kiyohara:2021smr}.
%We generate \textcolor{red}{$6\times10^5$} configurations separated by \textcolor{red}{five} sets of \textcolor{red}{five???} heat-bath updates followed by an over-relaxation step, at each of three simulation points $(\beta_{sim},\kappa_{sim})=???$ around the CP.
%\textcolor{red}{(KK: Please supplement concrete values of $(\beta_{sim},\kappa_{sim})$ as well as the number of configurations!)}
We then incorporate the next-to-leading order contributions of the HPE exactly using the reweighting method~\cite{Kiyohara:2021smr}. 
The effects of yet higher order contributions were shown to be small around the CP for $N_t=4$~\cite{Wakabayashi:2021eye}.
The CP of this system has been investigated using the Binder-cumulant method in Ref.~\cite{Kiyohara:2021smr}. 

\begin{figure}
    \centering
    % \begin{minipage}{0.35\linewidth}
    %     \includegraphics[width=\linewidth]{fig/HQQCD_Ratio-2.pdf}
    % \end{minipage}
    % \hspace{0.05\textwidth}
    % \begin{minipage}{0.35\linewidth}
    %     \includegraphics[width=\linewidth]{fig/HQQCD_Ratio-1.pdf}
    % \end{minipage}
    \includegraphics[width=0.7\linewidth]{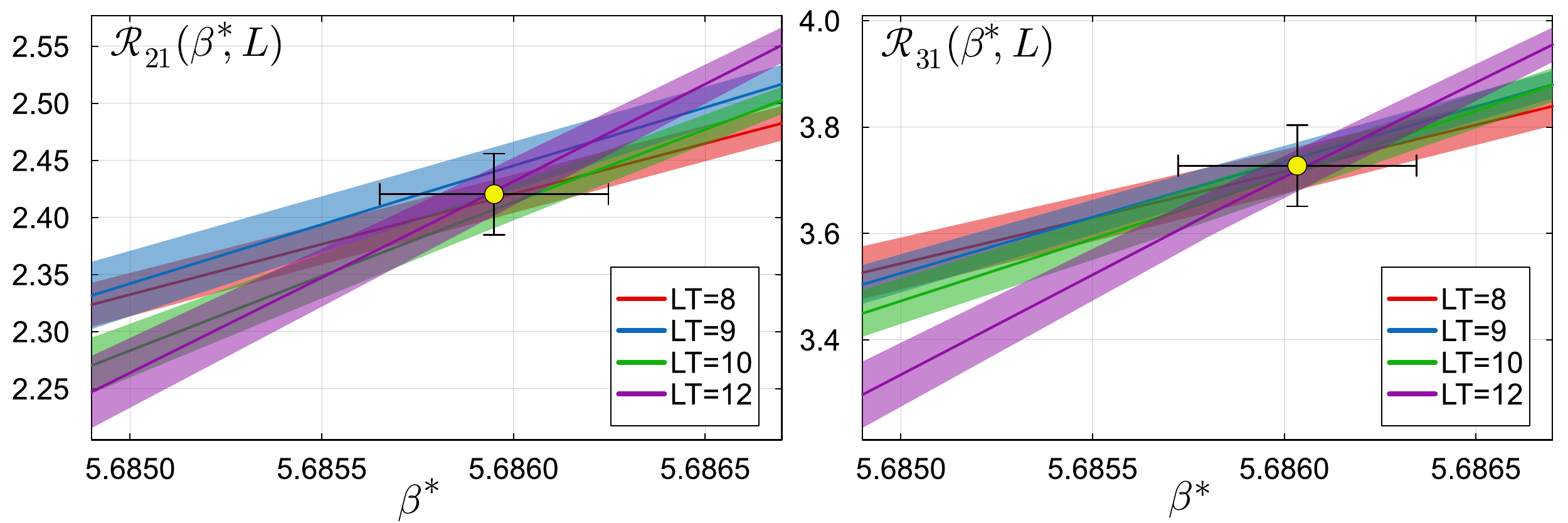}
    \caption{LYZR ${\cal R}_{21}(\tau,L)$ (left) and ${\cal R}_{31}(\tau,L)$ (right) in heavy-quark QCD at $N_t=4$ for various aspects ratios $LT=N_s/N_t$. The circles are the results of FSS fits.}
    \label{fig:HQ_R}
\end{figure}

Our results of the LYZR ${\cal R}_{21}$ and ${\cal R}_{31}$ are shown in
Fig.~\ref{fig:HQ_R} as functions of $\beta^*=\beta+16N_cN_f\kappa^4$, where the shaded bands represent statistical errors estimated by the jackknife method. % with bin-size 20. 
%Close to the CP, $\beta^*=\beta^*_\textrm{c}$, we can use the aspect ratio $LT$ as a parameter for the system size. 
We find that ${\cal R}_{n1}$ at various $LT$ are approximately crossing at a common intersection point.
Ignoring the second bracket of Eq.~(\ref{eq:Rlinear2}) for same reason as the Potts model, we perform three-parameter FSS fits ${\cal R}_{n1}(\beta^*,LT)=r+c(\beta^*-\beta^*_\textrm{c})(LT)^{y_t}$ for $LT\ge9$ ($12$ data points in total), where $r$, $c$, $\beta^*_\textrm{c}$, and $y_t$ are the fitting parameters.
% \textcolor{red}{(KK: Wada kun, I have replaced $L$ by $LT$. Accordingly, $L\ge40$ was replaced by $LT\ge10$. Please check if these are correct! I also want to confirm the meaning of ``12 data points''.)}
The open circles in Fig.~\ref{fig:HQ_R} show the fitting results $\beta^*_\textrm{c} = 5.68595(30)$ and $5.68603(31)$ for $\mathcal{R}_{21}$ and $\mathcal{R}_{31}$, respectably.
% \textcolor{red}{(KK: Wada kun, The circle for ${\cal R}_{21}$ looks strange if it is meant the intersection for $LT\ge10$. Please check if the fitting range is correct!)}
We find that these results are consistent with $\beta^*_c= 5.68578(22)$ obtained by a Binder cumulant analysis~\cite{Kiyohara:2021smr}.

\section{Summary}

In this study, we proposed a general method, the LYZR method, to locate the critical point (CP) from the numerical simulations performed on finite-size systems using the ratios of LYZ. 
This method is based on finite-size scaling of LYZ %so that the major finite-size effects on large systems are taken into account, 
similarly to the conventional Binder-cumulant method.
For studies in general systems, the LYZR method is superior to the Binder-cumulant method in faster suppression of the operator-mixing effects at large system sizes.
%The ratios of LYZ pass through one point at the CP whose value is characterized by the universality class. Consequently, LYZ ratios for different volumes intersect at the CP. 
To test the LYZR method, we performed Monte Carlo simulations in three models; the three-dimensional Ising model, the three-dimensional three-state Potts model, and finite-temperature QCD in the heavy-quark region, which have a CP in the same universality class. % of the Ising model. 
These results confirm the validity of the LYZR method in practical numerical analyses. 
% Applying this method to $2+1$-flavors QCD at physical-quark mass could yield more accurate results by accounting for finite-size effects. 
% However, the challenges remain: the second LYZ is far apart from the real axis and may hide for small space volume in the LYZ related to the Roberge-Weiss phase transition which appear at the $\textrm{Im}\mu_B=\pi/3$. To avoid confusion of the QCD-CP one with the Roberge-Weiss one, we should utilize the larger space volume lattice.
%The LYZR method should also be applicable to systems in universality classes other than the Ising model.
%Studies in this direction are also in progress.

\subsection*{Acknowledgements}
We thank Shinji Ejiri for discussions in the early stage of this study. The authors also thank Koji Hukushima, Attila P\'asztor and Christian Schmidt for useful discussions.
This work was supported in part by JSPS KAKENHI (Nos.~JP22K03593, JP22K03619, JP23H04507, JP24K07049), the Center for Gravitational Physics and Quantum Information (CGPQI) at Yukawa Institute for Theoretical Physics, the Research proposal-based use at the Cybermedia Center, Osaka University, and the Multidisciplinary Cooperative Research Program of the Center for Computational Sciences, University of Tsukuba.

\bibliographystyle{JHEP}
\bibliography{ref}

\providecommand{\href}[2]{#2}\begingroup\raggedright\begin{thebibliography}{10}

\bibitem{Wada:2024qsk}
T.~Wada, M.~Kitazawa and K.~Kanaya, \emph{{Lee-Yang-zero ratios for locating a critical point}},  \href{https://arxiv.org/abs/2410.19345}{{\ttfamily 2410.19345}}.

\bibitem{Kortman:1971zz}
P.J.~Kortman and R.B.~Griffiths, \emph{{Density of Zeros on the Lee-Yang Circle for Two Ising Ferromagnets}}, \href{https://doi.org/10.1103/PhysRevLett.27.1439}{\emph{Phys. Rev. Lett.} {\bfseries 27} (1971) 1439}.

\bibitem{Stephanov:2006dn}
M.A.~Stephanov, \emph{{QCD critical point and complex chemical potential singularities}}, \href{https://doi.org/10.1103/PhysRevD.73.094508}{\emph{Phys. Rev. D} {\bfseries 73} (2006) 094508} [\href{https://arxiv.org/abs/hep-lat/0603014}{{\ttfamily hep-lat/0603014}}].

\bibitem{Ejiri:2014oka}
S.~Ejiri, Y.~Shinno and H.~Yoneyama, \emph{{Complex singularities around the QCD critical point at finite densities}}, \href{https://doi.org/10.1093/ptep/ptu108}{\emph{PTEP} {\bfseries 2014} (2014) 083B02} [\href{https://arxiv.org/abs/1404.6004}{{\ttfamily 1404.6004}}].

\bibitem{An:2017brc}
X.~An, D.~Mesterh\'azy and M.A.~Stephanov, \emph{{On spinodal points and Lee-Yang edge singularities}}, \href{https://doi.org/10.1088/1742-5468/aaac4a}{\emph{J. Stat. Mech.} {\bfseries 1803} (2018) 033207} [\href{https://arxiv.org/abs/1707.06447}{{\ttfamily 1707.06447}}].

\bibitem{Basar:2021gyi}
G.~Basar, G.V.~Dunne and Z.~Yin, \emph{{Uniformizing Lee-Yang singularities}}, \href{https://doi.org/10.1103/PhysRevD.105.105002}{\emph{Phys. Rev. D} {\bfseries 105} (2022) 105002} [\href{https://arxiv.org/abs/2112.14269}{{\ttfamily 2112.14269}}].

\bibitem{Rennecke:2022ohx}
F.~Rennecke and V.V.~Skokov, \emph{{Universal location of Yang\textendash{}Lee edge singularity for a one-component field theory in 1\ensuremath{\leq}d\ensuremath{\leq}4}}, \href{https://doi.org/10.1016/j.aop.2022.169010}{\emph{Annals Phys.} {\bfseries 444} (2022) 169010} [\href{https://arxiv.org/abs/2203.16651}{{\ttfamily 2203.16651}}].

\bibitem{Johnson:2022cqv}
G.~Johnson, F.~Rennecke and V.V.~Skokov, \emph{{Universal location of Yang-Lee edge singularity in classic O(N) universality classes}}, \href{https://doi.org/10.1103/PhysRevD.107.116013}{\emph{Phys. Rev. D} {\bfseries 107} (2023) 116013} [\href{https://arxiv.org/abs/2211.00710}{{\ttfamily 2211.00710}}].

\bibitem{Singh:2023bog}
S.~Singh, M.~Cipressi and F.~Di~Renzo, \emph{{Exploring Lee-Yang and Fisher zeros in the 2D Ising model through multipoint Pad\'e approximants}}, \href{https://doi.org/10.1103/PhysRevD.109.074505}{\emph{Phys. Rev. D} {\bfseries 109} (2024) 074505} [\href{https://arxiv.org/abs/2312.03178}{{\ttfamily 2312.03178}}].

\bibitem{Karsch:2023rfb}
F.~Karsch, C.~Schmidt and S.~Singh, \emph{{Lee-Yang and Langer edge singularities from analytic continuation of scaling functions}}, \href{https://doi.org/10.1103/PhysRevD.109.014508}{\emph{Phys. Rev. D} {\bfseries 109} (2024) 014508} [\href{https://arxiv.org/abs/2311.13530}{{\ttfamily 2311.13530}}].

\bibitem{Dimopoulos:2021vrk}
P.~Dimopoulos, L.~Dini, F.~Di~Renzo, J.~Goswami, G.~Nicotra, C.~Schmidt et~al., \emph{{Contribution to understanding the phase structure of strong interaction matter: Lee-Yang edge singularities from lattice QCD}}, \href{https://doi.org/10.1103/PhysRevD.105.034513}{\emph{Phys. Rev. D} {\bfseries 105} (2022) 034513} [\href{https://arxiv.org/abs/2110.15933}{{\ttfamily 2110.15933}}].

\bibitem{Basar:2023nkp}
G.~Basar, \emph{{On the QCD critical point, Lee-Yang edge singularities and Pade resummations}},  \href{https://arxiv.org/abs/2312.06952}{{\ttfamily 2312.06952}}.

\bibitem{Zambello:2023ptp}
K.~Zambello, D.A.~Clarke, P.~Dimopoulos, F.~Di~Renzo, J.~Goswami, G.~Nicotra et~al., \emph{{Determination of Lee-Yang edge singularities in QCD by rational approximations}}, \href{https://doi.org/10.22323/1.430.0164}{\emph{PoS} {\bfseries LATTICE2022} (2023) 164} [\href{https://arxiv.org/abs/2301.03952}{{\ttfamily 2301.03952}}].

\bibitem{Clarke:2024ugt}
D.A.~Clarke, P.~Dimopoulos, F.~Di~Renzo, J.~Goswami, C.~Schmidt, S.~Singh et~al., \emph{{Searching for the QCD critical endpoint using multi-point Pad\'e approximations}},  \href{https://arxiv.org/abs/2405.10196}{{\ttfamily 2405.10196}}.

\bibitem{Skokov:2024fac}
V.V.~Skokov, \emph{{Two lectures on Yang-Lee edge singularity and analytic structure of QCD equation of state}},  \href{https://arxiv.org/abs/2411.02663}{{\ttfamily 2411.02663}}.

\bibitem{Binder:1981sa}
K.~Binder, \emph{{Finite size scaling analysis of Ising model block distribution functions}}, \href{https://doi.org/10.1007/BF01293604}{\emph{Z. Phys. B} {\bfseries 43} (1981) 119}.

\bibitem{Ferrenberg:2018zst}
A.M.~Ferrenberg, J.~Xu and D.P.~Landau, \emph{{Pushing the limits of Monte Carlo simulations for the three-dimensional Ising model}}, \href{https://doi.org/10.1103/PhysRevE.97.043301}{\emph{Phys. Rev. E} {\bfseries 97} (2018) 043301} [\href{https://arxiv.org/abs/1806.03558}{{\ttfamily 1806.03558}}].

\bibitem{Lee:1952ig}
T.D.~Lee and C.-N.~Yang, \emph{{Statistical theory of equations of state and phase transitions. 2. Lattice gas and Ising model}}, \href{https://doi.org/10.1103/PhysRev.87.410}{\emph{Phys. Rev.} {\bfseries 87} (1952) 410}.

\bibitem{Ejiri:2005ts}
S.~Ejiri, \emph{{Lee-Yang zero analysis for the study of QCD phase structure}}, \href{https://doi.org/10.1103/PhysRevD.73.054502}{\emph{Phys. Rev. D} {\bfseries 73} (2006) 054502} [\href{https://arxiv.org/abs/hep-lat/0506023}{{\ttfamily hep-lat/0506023}}].

\bibitem{BENA_2005}
I.~Bena, M.~Droz and A.~LIPOWSKI, \emph{Statistical mechanics of equilibrium and nonequilibrium phase transitions: The yang–lee formalism}, \href{https://doi.org/10.1142/s0217979205032759}{\emph{International Journal of Modern Physics B} {\bfseries 19} (2005) 4269–4329}.

\bibitem{Wolff:1988uh}
U.~Wolff, \emph{{Collective Monte Carlo Updating for Spin Systems}}, \href{https://doi.org/10.1103/PhysRevLett.62.361}{\emph{Phys. Rev. Lett.} {\bfseries 62} (1989) 361}.

\bibitem{PhysRevLett.61.2635}
A.M.~Ferrenberg and R.H.~Swendsen, \emph{New monte carlo technique for studying phase transitions}, \href{https://doi.org/10.1103/PhysRevLett.61.2635}{\emph{Phys. Rev. Lett.} {\bfseries 61} (1988) 2635}.

\bibitem{Kos:2016ysd}
F.~Kos, D.~Poland, D.~Simmons-Duffin and A.~Vichi, \emph{{Precision Islands in the Ising and $O(N)$ Models}}, \href{https://doi.org/10.1007/JHEP08(2016)036}{\emph{JHEP} {\bfseries 08} (2016) 036} [\href{https://arxiv.org/abs/1603.04436}{{\ttfamily 1603.04436}}].

\bibitem{Wada:2025xx}
T.~Wada, M.~Kitazawa and K.~Kanaya, \emph{{in preparation}}, .

\bibitem{Rehr:1973zz}
J.J.~Rehr and N.D.~Mermin, \emph{{Revised Scaling Equation of State at the Liquid-Vapor Critical Point}}, \href{https://doi.org/10.1103/PhysRevA.8.472}{\emph{Phys. Rev. A} {\bfseries 8} (1973) 472}.

\bibitem{Wilding:1997xx}
N.B.~Wilding, \emph{{Simulation studies of fluid critical behaviour}}, \href{https://doi.org/10.1088/0953-8984/9/3/002}{\emph{J. Phys.: Condens. Matter} {\bfseries 9} (1997) 585}.

\bibitem{Karsch:2000xv}
F.~Karsch and S.~Stickan, \emph{{The Three-dimensional, three state Potts model in an external field}}, \href{https://doi.org/10.1016/S0370-2693(00)00902-3}{\emph{Phys. Lett. B} {\bfseries 488} (2000) 319} [\href{https://arxiv.org/abs/hep-lat/0007019}{{\ttfamily hep-lat/0007019}}].

\bibitem{Cuteri:2020yke}
F.~Cuteri, O.~Philipsen, A.~Sch\"on and A.~Sciarra, \emph{{Deconfinement critical point of lattice QCD with $N_f$=2 Wilson fermions}}, \href{https://doi.org/10.1103/PhysRevD.103.014513}{\emph{Phys. Rev. D} {\bfseries 103} (2021) 014513} [\href{https://arxiv.org/abs/2009.14033}{{\ttfamily 2009.14033}}].

\bibitem{Kiyohara:2021smr}
A.~Kiyohara, M.~Kitazawa, S.~Ejiri and K.~Kanaya, \emph{{Finite-size scaling around the critical point in the heavy quark region of QCD}}, \href{https://doi.org/10.1103/PhysRevD.104.114509}{\emph{Phys. Rev. D} {\bfseries 104} (2021) 114509} [\href{https://arxiv.org/abs/2108.00118}{{\ttfamily 2108.00118}}].

\bibitem{Ashikawa:2024njc}
R.~Ashikawa, M.~Kitazawa, S.~Ejiri and K.~Kanaya, \emph{{High-precision analysis of the critical point in heavy-quark QCD at Nt=6}}, \href{https://doi.org/10.1103/PhysRevD.110.074508}{\emph{Phys. Rev. D} {\bfseries 110} (2024) 074508} [\href{https://arxiv.org/abs/2407.09156}{{\ttfamily 2407.09156}}].

\bibitem{Wakabayashi:2021eye}
N.~Wakabayashi, S.~Ejiri, K.~Kanaya and M.~Kitazawa, \emph{{Scope and convergence of the hopping parameter expansion in finite-temperature quantum chromodynamics with heavy quarks around the critical point}}, \href{https://doi.org/10.1093/ptep/ptac019}{\emph{PTEP} {\bfseries 2022} (2022) 033B05} [\href{https://arxiv.org/abs/2112.06340}{{\ttfamily 2112.06340}}].

\end{thebibliography}\endgroup

\end{document}